# From Text to Map: A System Dynamics Bot for Constructing Causal Loop Diagrams

Niyousha Hosseinichimeh, Aritra Majumdar, Ross Williams, Navid Ghaffarzadegan

Department of Industrial and Systems Engineering, Virginia Tech, Falls Church, VA

Contact: niyousha@vt.edu

**Abstract**

We introduce and test the System Dynamics Bot, a computer program leveraging a large language model to automate the creation of causal loop diagrams from textual data. To evaluate its performance, we ensembled two distinct databases. The first dataset includes 20 causal loop diagrams and associated texts sourced from the system dynamics literature. The second dataset comprises responses from 30 participants to a vignette, along with causal loop diagrams coded by three system dynamics modelers. The bot uses textual data and successfully identifies approximately sixty percent of the links between variables and feedback loops in both datasets. This paper outlines our approach, provides examples, and presents evaluation results. We discuss encountered challenges and implemented solutions in developing the System Dynamics Bot. The bot can facilitate extracting mental models from textual data and improve model building processes. Moreover, the two datasets can serve as a testbed for similar programs.

## 1. Introduction

Constructing causal loop diagrams (CLDs) is a foundational step in analyzing complex systems and building systems models (Sterman, 2000). CLDs have been integral to system dynamics since its inception and have demonstrated effectiveness in engaging stakeholders, sharing their mental models, and facilitating the collaborative discussions to refine and enhance them (Baugh Littlejohns, Baum, Lawless, & Freeman, 2018; Black, 2013; Black & Andersen, 2012). The quality of such CLDs, therefore, is important since misalignment between the objective reality and subjective mental models often is a barrier for effective decision-making (Kim & Andersen, 2012).

Systematic methods have been established for developing CLDs from textual data. Kim and Andersen (2012) proposed a coding approach for developing CLDs based on the grounded theory, involving distinct steps of discovering themes in the data, identifying variables and their causal relationships, transferring text into words-and-arrow diagrams, generalizing the structural representations, and retaining links between the maps and the data source (Kim & Andersen, 2012). While Kim and Andersen's method provides a rigorous and systematic approach for constructing CLDs from text data, it is labor intensive. To overcome this limitation, alternative approaches have been developed, modifying Kim and Andersen's original method (Eker & Zimmermann, 2016; Turner, Kim, & Andersen, 2013). While different software packages have been used to better document and retain the links between the maps and the data sources (Yearworth & White, 2013), within the system dynamics community, none of the past approaches for constructing the maps are fully automatic.

Our objective is to build on a series of efforts from computer science to extract causal argument from text with main applications for system dynamics. In this paper, we report on our research effort to use generative artificial intelligence (AI) to automate developing CLDs from textual data. More specifically, we developed a computer program, System Dynamics Bot (SD Bot), powered by GPT-4-Turbo to identify causality in a text, identify relationships, merge them, and develop a CLD. We also compiled two datasets to evaluate such computer programs. In the following, we discuss the background of efforts to extract causal relationship from text, and the new advancements in generative AI as related to system dynamics. We then introduce the SD Bot, and evaluate its performance.

## 2. Background

Beyond the domain of system dynamics, the automatic extraction of information about causality from text have been explored widely over the past decades (Chan & Lam, 2005; Kaplan & Berry-Rogghe, 1991; Khoo, Kornfilt, Oddy, & Myaeng, 1998; Yang, Han, & Poon, 2022). A recent survey of systems developed to identify causal relationship from text categorized them into three main groups of 1) knowledge-based approaches, 2) statistical machine learning-based approaches, and 3) deep learning-based approaches (Yang et al 2022).

Knowledge based approaches use patterns such as causal verbs (Garcia, EDF-DER, & IMA-TIEM, 1997) or expressions (Khoo et al., 1998) to identify causal relationships. Statistical machine learning-based approaches typically utilize third-party natural language processing tools to generate features for labeled data, and subsequently, machine learning algorithms like support vector machine, naive Bayes, and logistic regression are employed for classification tasks (Lin, Kan, & Ng, 2009). Deep learning is a branch of machine learning inspired by how human brain works. Relative to the first two approaches, deep learning excels at automatically extracting higher-level information from input vectors and adjusting expected results, allowing these systems to prioritize input features and model architecture over linguistic information preparation (Yang et al., 2022). However, deep learning-based systems demand access to larger datasets and more significant computational resources compared to other techniques.

These systems are developed to identify cause-and-effect pairs, and not necessarily a CLD. In this paper, our focus is on a more complex task of extracting CLDs, as a major step in developing system dynamics models. The new opportunity that can make automating the practice of building system dynamics models possible is related to recent improvements in generative AI. Recent studies have shown that generative AI can be used for building simulation environments.

(Park et al., 2023; Park et al., 2022; Wang et al., 2023) and system dynamics models (Ghaffarzadegan, Majumdar, Williams, & Hosseinichimeh, 2024; Williams, Hosseinichimeh, Majumdar, & Ghaffarzadegan, 2023). While the field of building simulation models with generative AI is expanding (Argyle et al., 2023; Dillion, Tandon, Gu, & Gray, 2023; Hamilton, 2023; Horton, 2023), the contributions of the new technology to system dynamics can go beyond simulating human behavior (Akhavan & Jalali, 2023). Our focus is on using generative AI for the purpose of developing CLDs.

In this paper, we present a new approach for developing a CLD from text that uses a large language model (LLM), specifically GPT 4, to identify causality and map the causal relationships. Instead of using conventional methods of training a deep learning model on a large dataset, we leverage the LLM's inherent understanding of text and combine it with few-shot prompting techniques to train our SD Bot. To assess its performance, we conduct two types of experiments. The following sections explain our SD Bot and describe how it is evaluated.

### 3. Our solution: SD Bot

The SD Bot receives a piece of text as an input and reports the identified causal relationships, reasoning associated with each identified relationship, and the relevant section of the text where the relationships are extracted. Then the bot creates a CLD. The program is developed in Python and is trained using few-shot prompting technique in JSON format. These prompts are accompanied by step-by-step instructions to identify relationships between variables, considering both positive and negative connections. Upon analyzing the text and creating a set of causal links between different variables and the polarity of the relationships, the system parses it, treating each variable name as a node in a network. A Python module, networkx, is utilized to convert the resulting interconnected network of variables and relationships into an image, which can be optionally provided as an additional form of output. The program uses generative AI for this process, and specifically uses GPT-4-Turbo (GPT-4-1106-preview) as its large language model.

For the purpose of introduction, our first example is the output of the bot to a simple text: "More chickens lay more eggs, which hatch and add to the chicken population. The more chickens, the more road crossing they will attempt. If there is any traffic, more road crossings will lead to fewer chickens" (Sterman 2000, p.13). Figure 1 shows the results. Panel a reports the text input, and the identified causal relationships, polarity of the causal links, reasoning associated with each relationship, and the relevant section of the text. As shown, causal links with proper polarities are identified. Each causal relationship is linked to a specific part of the text that implies the relationship. Using these identified causes, then a CLD is developed. Panel b compares the CLD generated by the SD Bot (left) and the CLD presented in the source (right).

**SD Bot input:**
*"More chickens lay more eggs, which hatch and add to the chicken population. The more chickens, the more road crossing they will attempt. If there is any traffic, more road crossings will lead to fewer chickens."*

**SD Bot output:**
1. chickens --> (+) eggs
Reasoning: The relationship indicates that an increase in the number of chickens leads to an increase in egg production.
Relevant Text: More chickens lay more eggs, which hatch and add to the chicken population.

2. eggs --> (+) chickens
Reasoning: This relationship shows that an increase in eggs results in an increase in the chicken population, as eggs hatch into chickens.
Relevant Text: More chickens lay more eggs, which hatch and add to the chicken population.

3. chickens --> (+) road crossings
Reasoning: As the number of chickens increases, the frequency of road crossings attempted by chickens also increases.
Relevant Text: The more chickens, the more road crossing they will attempt.

4. road crossings --> (-) chickens
Reasoning: This relationship suggests that an increase in road crossings, due to traffic, results in a decrease in the number of chickens, likely due to accidents.
Relevant Text: If there is any traffic, more road crossings will lead to fewer chickens.

a.

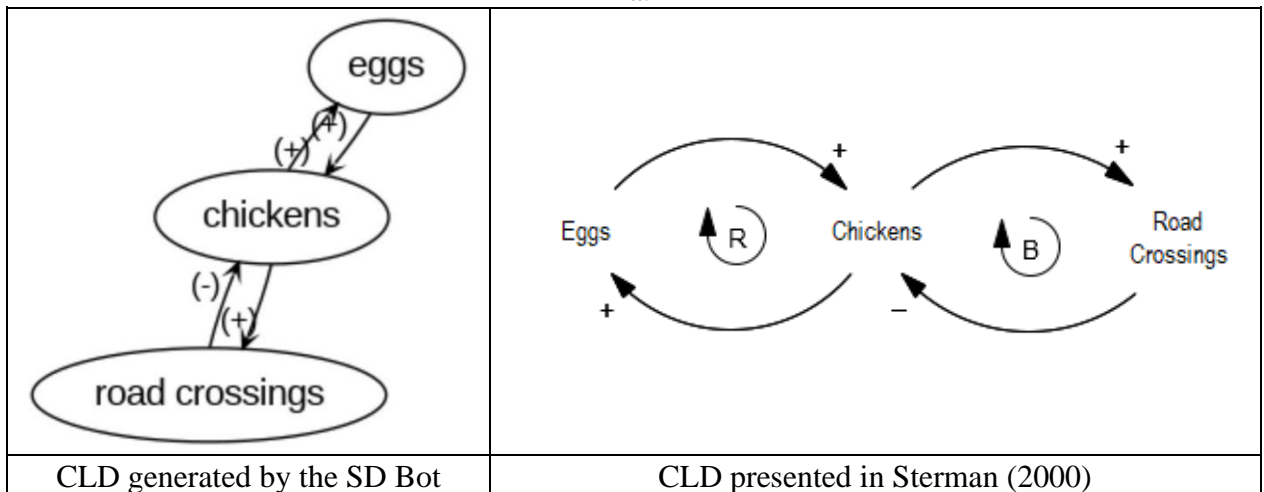

| CLD generated by the SD Bot | CLD presented in Sterman (2000) |

b.

Figure 1. An example of the SD Bot's outcome: a) The causal relationships, reasoning, and relevant text offered by the SD Bot in response to the input; b) the generative CLD (left) versus the ground truth from the source (right).

While effective for simple texts, challenges arise in more complex cases, such as text inputs containing different variable names for the same concept, misinterpreting relationships between variables due to existence of adjectives before variables, and incomplete loops. Further elaborations on each of these challenges and how we address them follow. For example, people usually use similar words and phrases for the same concepts. To address instances of similar variables, a similarity matrix of size $m \times m$, where m is the number of variables in the revised response, is generated. Similarities are determined by representing each variable as an embedding and calculating cosine similarities between each unique pair of variables. By applying a threshold value to the scores, we filter and identify pairs of variables with the highest similarities. Users are then prompted to confirm or select groups of similar variables for merging. The interaction includes asking the user to decide whether to retain all variables, specific groups, or proceed with merging individual groups. The merging process entails updating causal relationships based on the merged variables.

After obtaining the merged response, we check the polarities of the relationships. We examine identified causal relationships in the form of variable X influencing variable Y, against four different conditions related to the text:

Condition 1: Increasing X increases Y.

Condition 2: Decreasing X decreases Y.

Condition 3: Increasing X decreases Y.

Condition 4: Decreasing X increases Y.

This process helps assure that polarities are correctly identified. If conditions 1 and 2 are true, the polarity of the link is positive, and if conditions 3 and 4 are satisfied, the polarity is negative. The relationships are then edited accordingly, and in the majority of cases, accurate relationships between variables are identified.

When variables have adjectives, the relationships might not be captured correctly. For example, before adding reasoning, the earlier version of the bot mistakenly identified a negative relationship between product attractiveness and order rate in the following text, "Orders booked increase the order backlog which increases the delivery delay which makes the product less attractive and reduces the order rate (Forrester, 1968)." To mitigate misinterpretations of relationships, we incorporated the pertinent line from the text, which served as the basis for extracting the causal relationship, along with the reasoning provided by the SD Bot for each identified relationship. These served as reference during the assessment of relationship polarity against the four conditions outlined above. For the specified example, the associated reasoning generated by the bot is, "Reduced product attractiveness leads to a decrease in the order rate." The SD Bot checks the relationship with the reasoning and relevant line as references against the above four conditions and corrects it.

To handle cases of incomplete loops, a simple follow-up question is posed, examining whether there are any implied possibilities of forming closed loops in the text. If so, the loops are closed by adding additional relationships. This instructs the bot to reassess its initial response, allowing the identification and resolution of any existing loops in the text, with the revised response incorporating the identified loops.

Figure 2 shows another example of SD Bot's outcome. Panel a illustrates the text input (Forrester, 1968) and the bot's output, including the causal relationships with polarity of the causal links, reasoning, and related text. Panel b compares the CLD generated by the SD Bot (left) and the CLD presented in the article (right). The SD Bot correctly captures the reinforcing loop that depicts the growth of the firm: more orders booked lead to more revenue, more sales budget, and more salesmen which further increases orders booked. Additionally, it captures two balancing feedback loops: one representing the decline in orders booked due to increased delivery delays, and the other illustrating the increase in production capacity with a delay following a rise in delivery delays. In addition to identifying a causal path from delivery delay to production capacity via capacity order, the SD Bot has also included a redundant link that connects delivery delay directly to the production capacity. This creates a redundant balancing feedback loop.

**SD Bot input:**
*"Here is a situation in which salesmen book orders followed by product delivery which generates revenue which produces the sales budget which permits hiring still more salesmen. In short, salesmen produce revenue to pay for the further expansion of the sales effort. Orders booked increase the order backlog which increases the delivery delay which makes the product less attractive and reduces the order rate. The ordering of new production capacity is a function of delivery delay only. Rising order backlog, as indicated by delivery delay, is taken as an indication of inadequate capacity, and orders for more capacity are placed. These orders, after an acquisition delay, add to the production capacity. As the delivery delay rises, production capacity is raised to bring down the delivery delay."*

**SD Bot output:**
1. Salesmen --> (+) Orders booked
Reasoning: Salesmen are responsible for booking orders which lead to revenue generation.
Relevant Text: Here is a situation in which salesmen book orders followed by product delivery which generates revenue which produces the sales budget which permits hiring still more salesmen.

2. Revenue --> (+) Sales budget
Reasoning: Revenue generated from orders booked contributes to the sales budget.
Relevant Text: Here is a situation in which salesmen book orders followed by product delivery which generates revenue which produces the sales budget which permits hiring still more salesmen.

3. Sales budget --> (+) Salesmen
Reasoning: The sales budget allows for the hiring of more salesmen, expanding the sales effort.
Relevant Text: Here is a situation in which salesmen book orders followed by product delivery which generates revenue which produces the sales budget which permits hiring still more salesmen.

4. Orders booked --> (+) Order backlog
Reasoning: Orders booked contribute to the order backlog.
Relevant Text: Orders booked increase the order backlog which increases the delivery delay which makes the product less attractive and reduces the order rate.

5. Order backlog --> (+) Delivery delay
Reasoning: An increase in order backlog leads to longer delivery delays.
Relevant Text: As the delivery delay rises, production capacity is raised to bring down the delivery delay.

6. Delivery delay --> (-) Product attractiveness
Reasoning: Longer delivery delays make the product less attractive to customers.
Relevant Text: Orders booked increase the order backlog which increases the delivery delay which makes the product less attractive and reduces the order rate.

**SD Bot output-continued:**

7. Product attractiveness --> (-) Order rate
Reasoning: Reduced product attractiveness leads to a decrease in the order rate.
Relevant Text: Orders booked increase the order backlog which increases the delivery delay which makes the product less attractive and reduces the order rate.

8. Delivery delay --> (+) Capacity orders
Reasoning: The decision to order new production capacity is based on the current delivery delay.
Relevant Text: The ordering of new production capacity is a function of delivery delay only.

9. Delivery delay --> (+) Capacity orders
Reasoning: Delivery delay is used as an indicator of inadequate capacity, prompting orders for more capacity.
Relevant Text: Rising order backlog, as indicated by delivery delay, is taken as an indication of inadequate capacity, and orders for more capacity are placed.

10. Capacity orders --> (+) Production capacity
Reasoning: After an acquisition delay, the ordered capacity adds to the existing production capacity.
Relevant Text: These orders, after an acquisition delay, add to the production capacity.

11. Delivery delay --> (+) Production capacity
Reasoning: As delivery delay increases, production capacity is raised to reduce the delivery delay.
Relevant Text: As the delivery delay rises, production capacity is raised to bring down the delivery delay.

12. Production capacity --> (-) Delivery delay
Reasoning: Increased production capacity should reduce the delivery delay.
Relevant Text: As the delivery delay rises, production capacity is raised to bring down the delivery delay.

13. Order rate --> (-) Orders booked
Reasoning: Decreased order rate leads to fewer orders booked.
Relevant Text: Orders booked increase the order backlog which increases the delivery delay which makes the product less attractive and reduces the order rate.

14. Orders booked --> (-) Revenue
Reasoning: Fewer orders booked would lead to a decrease in revenue.
Relevant Text: Here is a situation in which salesmen book orders followed by product delivery which generates revenue which produces the sales budget which permits hiring still more salesmen.

a

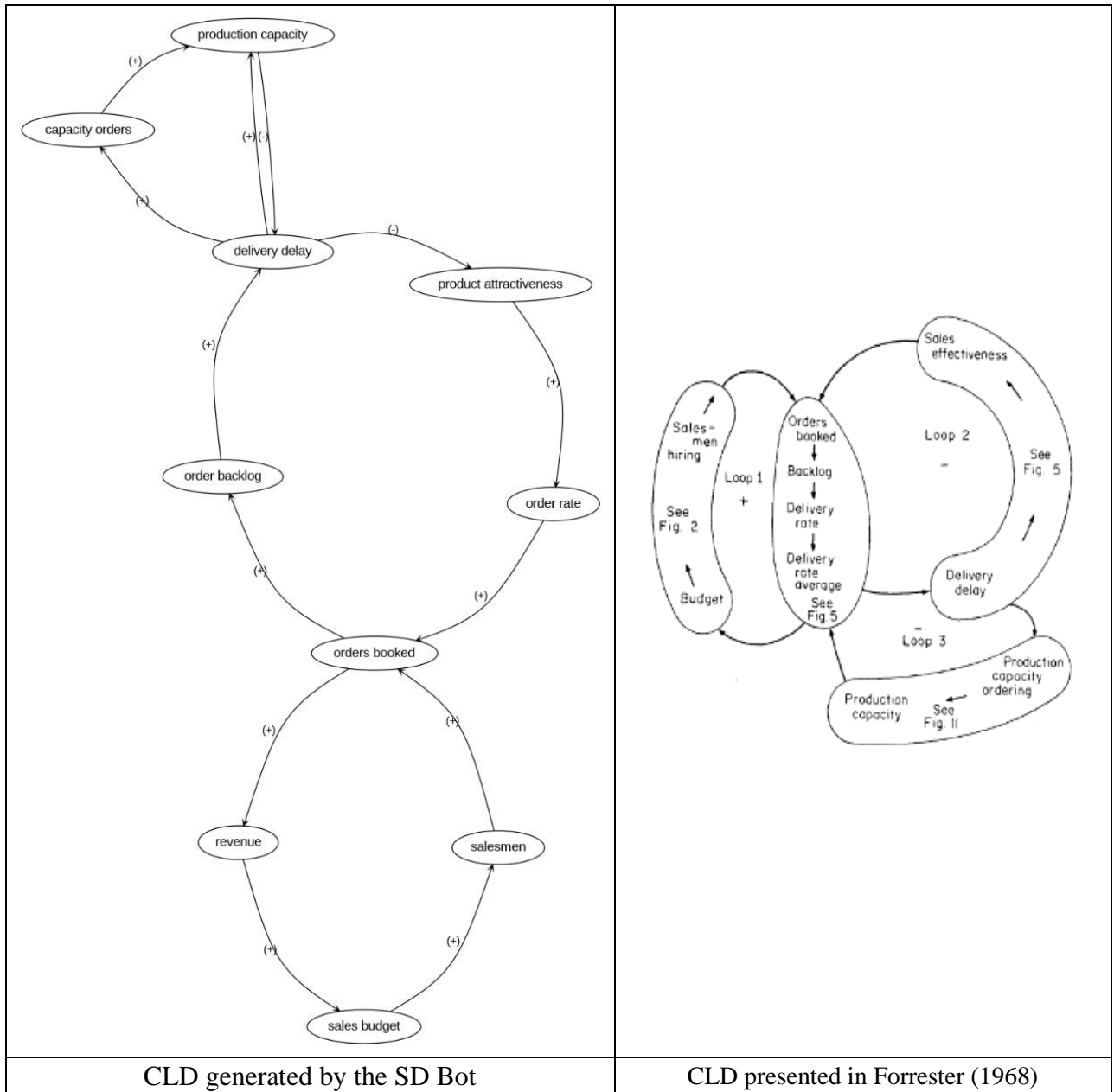

| CLD generated by the SD Bot | CLD presented in Forrester (1968) |

b

Figure 2. An example of SD Bot's outcome for a longer text: a) The causal relationships, reasoning, and relevant text offered by the SD Bot in response to the input; b) the generative CLD (left) versus the ground truth from the source (right).

In the next section, we systematically evaluate the performance of the SD Bot in different tests.

## 4. SD Bot evaluation

### 4.1. Method

We use two distinct datasets to systematically evaluate the SD Bot. The first dataset is compiled by gathering CLDs and corresponding texts from system dynamics textbooks and journal articles authored by system dynamics modelers. Specifically, we looked for relatively short pieces of text describing a CLD in system dynamics publications. This dataset comprises 20 examples of pairs of text and their associated CLDs. The performance of the SD Bot is then evaluated by comparing the CLD generated by the SD Bot from the text against the CLDs presented in the sources (ground truth). These examples vary in terms of text size and clarity of the text that describes the relationships; thus, we do not expect a perfect performance. The largest text has 470 words and the smallest one includes 19 words. Some texts do not provide enough details about the CLD or the authors do not explain all pieces of a feedback loop, while some texts fully explain the loops. The mix helps to assess the performance of the bot in a variety of the situations.

In order to use the texts as inputs, we drop any mention or use of the term "feedback loop" to make sure SD Bot does not capture the loop because the text indicated the term. Also, if the CLD is a generic model (such as the capability trap model) and the text includes examples for describing each generic variable, we drop the examples to keep the text short and focused on the CLD.

To evaluate the SD Bot, we feed the text to the SD Bot, generate the CLD, and then count the number of links and loops shown in the references that the SD Bot captures correctly. The examples that are used to train the program are not used in the evaluation dataset. The supplementary material presents the references with associated texts and CLDs.

The second dataset comes from a published study and includes responses of 30 participants to a vignette describing a complex socio-environmental problem (Davis et al., 2020). These participants are not system dynamics experts and many of them do not think in terms of feedback loops. This test helps us to expand our evaluations to depicting mental models of non-experts. The text responses from these participants were individually analyzed through independent coding by three system dynamics modelers, who noted the number of links and loops (if any) in each response. Any discrepancies were resolved through collaborative case reviews. Then they used the codes to manually develop causal maps for each individual response (Haque, Mahmoudi, Ghaffarzadegan, & Triantis, 2023).

The manually coded maps by the system dynamics experts are assumed as the ground truth in this test and are used to evaluate SD Bot's performance. Specifically, in this test, we feed the text provided by each participant to the SD Bot, generate a CLD, and then compare the links and loops generated by the SD Bot with those coded by SD expert human coders. The supplementary material shows the participants' responses and associated CLDs generated by the SD modelers. The two databases can be served as a testbed for future bots.

### 4.2. Results

**Experiment 1**

Table 1 shows the results of our first experiment. On average, the SD Bot identifies 59% of the links and 66% of the loops presented in the references. The median percentage of links and loops presented in the references that are captured by the SD Bot is 58%. We sorted the cases by the number of links in Table 1. The number of links in the reference texts may represent the complexity of the case.

Table 1. Comparison of links and loops in Experiment 1

| Case | Reference (Ground truth) | | SD Bot | | Link match | Feedback loop match |
|---|---|---|---|---|---|---|
| | Link | Loops | Links | Loops | | |
| 1 | 4 | 2 | 4 | 2 | 100% | 100% |
| 2 | 5 | 2 | 4 | 2 | 80% | 100% |
| 3 | 6 | 1 | 5 | 1 | 83% | 100% |
| 4 | 6 | 3 | 4 | 1 | 67% | 33% |
| 5 | 6 | 3 | 4 | 2 | 67% | 67% |
| 6 | 8 | 1 | 2 | 1 | 25% | 100% |
| 7 | 8 | 2 | 6 | 2 | 75% | 100% |
| 8 | 9 | 2 | 6 | 1 | 67% | 50% |
| 9 | 9 | 2 | 4 | 1 | 44% | 50% |
| 10 | 9 | 1 | 4 | 1 | 44% | 100% |
| 11 | 11 | 4 | 6 | 2 | 55% | 50% |
| 12 | 11 | 4 | 6 | 1 | 55% | 25% |
| 13 | 12 | 3 | 12 | 3 | 100% | 100% |
| 14 | 17 | 3 | 11 | 2 | 65% | 67% |
| 15 | 17 | 6 | 4 | 2 | 24% | 33% |
| 16 | 18 | 4 | 11 | 3 | 61% | 75% |
| 17 | 21 | 6 | 9 | 3 | 43% | 50% |
| 18 | 22 | 3 | 7 | 1 | 32% | 33% |
| 19 | 23 | 6 | 10 | 3 | 43% | 50% |
| 20 | 23 | 9 | 11 | 3 | 48% | 33% |
| Mean | | | | | 59% | 66% |
| Median | | | | | 58% | 58% |

Figure 3 shows the distribution of percent links and feedback loops correctly identified by the SD Bot presented in Table 1. In half of the cases, 70% or more of the links were correctly identified (panel a). Additionally, in half of the cases, 70% or more of the feedback loops were correctly identified. In 35% of the cases, all of the feedback loops were identified correctly (7 out of 20) (panel b).

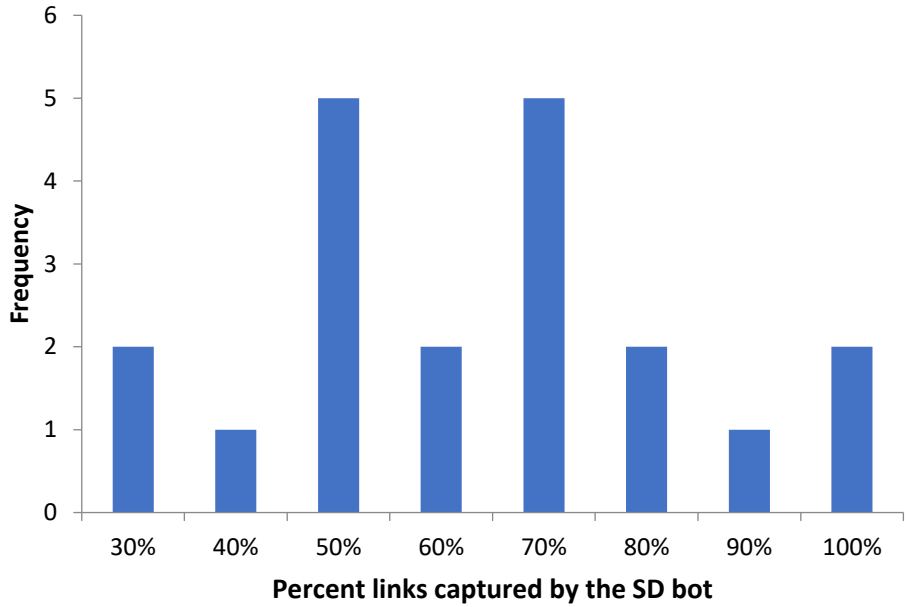

a. Frequency of percent links in references captured by the SD Bot

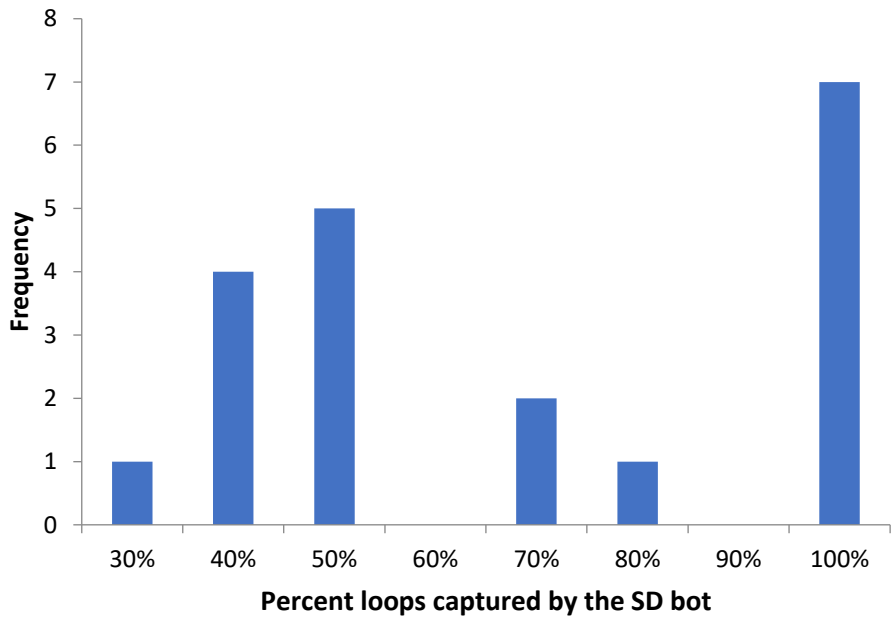

b. Frequency of percent loops in references captured by the SD Bot
Figure 3. Distribution of percent links and loops captured by the SD Bot in experiment 1

**Experiment 2**

In the second experiment, we tested the extent to which the SD Bot can capture the mental model of participants who completed a vignette. On average, the bot captured 56% of the links. In addition, we compared whether human coders and the bot agree on the existence of feedback loops in responses. In 83% of cases the loop results match (Table 2).

Table 2. Comparison of links and loops in Experiment 2

| Participants | Human coders (Ground truth) | | SD Bot | | Link match | Feedback loop match |
|---|---|---|---|---|---|---|
| | Link | Loops | Links | Loops | | |
| 1 | 1 | 0 | 1 | 0 | 100% | Yes |
| 2 | 1 | 0 | 0 | 0 | 0% | Yes |
| 3 | 2 | 0 | 2 | 0 | 100% | Yes |
| 4 | 2 | 0 | 0 | 0 | 0% | Yes |
| 5 | 3 | 0 | 2 | 1 | 67% | No |
| 6 | 4 | 1 | 3 | 1 | 75% | Yes |
| 7 | 4 | 0 | 4 | 0 | 100% | Yes |
| 8 | 4 | 0 | 0 | 0 | 0% | Yes |
| 9 | 5 | 0 | 4 | 0 | 80% | Yes |
| 10 | 5 | 0 | 5 | 0 | 100% | Yes |
| 11 | 5 | 0 | 3 | 0 | 60% | Yes |
| 12 | 5 | 0 | 4 | 0 | 80% | Yes |
| 13 | 5 | 0 | 3 | 0 | 60% | Yes |
| 14 | 6 | 0 | 4 | 0 | 67% | Yes |
| 15 | 5 | 0 | 3 | 0 | 60% | Yes |
| 16 | 6 | 1 | 0 | 0 | 0% | No |
| 17 | 6 | 0 | 4 | 0 | 67% | Yes |
| 18 | 7 | 0 | 5 | 0 | 71% | Yes |
| 19 | 7 | 0 | 4 | 0 | 57% | Yes |
| 20 | 7 | 0 | 2 | 0 | 29% | Yes |
| 21 | 10 | 0 | 6 | 0 | 60% | Yes |
| 22 | 8 | 0 | 4 | 0 | 50% | Yes |
| 23 | 8 | 0 | 2 | 1 | 25% | No |
| 24 | 9 | 1 | 5 | 0 | 56% | No |
| 25 | 9 | 0 | 8 | 0 | 89% | Yes |
| 26 | 9 | 1 | 0 | 0 | 0% | No |
| 27 | 11 | 0 | 7 | 0 | 64% | Yes |
| 28 | 11 | 0 | 4 | 0 | 36% | Yes |
| 29 | 14 | 4 | 8 | 3 | 57% | Yes (75.0%) |
| 30 | 14 | 0 | 9 | 0 | 64% | Yes |
| Mean | | | | | 56% | 83% |
| Median | | | | | 60% | 100% |

Figure 4 presents the frequency of the links captured by the SD Bot. In 21 cases (out of 30), 60% or more of the links were correctly identified by the SD Bot. There are five cases in which the bot did not identify any causality between variables in the participants' responses (Figure 4).

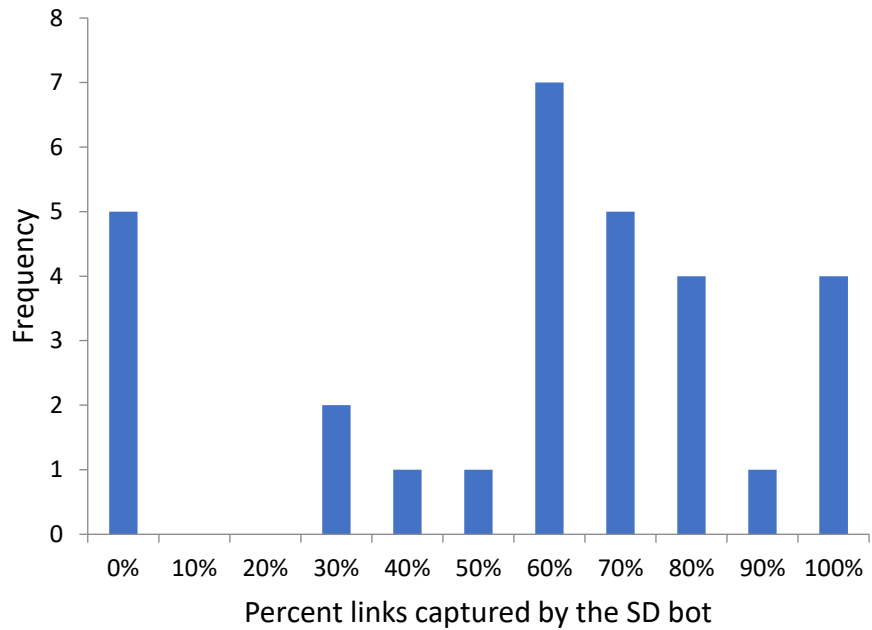

Figure 4. Distribution of % linked captured by the SD Bot in experiment 2

In 25 cases out of 30, the identified feedback loops by the SD Bot match with the SD modelers' results. It should be noted that the response of the majority of participants did not include any feedback loop and the SD Bot correctly identified zero loops. This confirms a high rate of True Negatives which is crucial for using the bot for mapping non-SD experts' mental models. As shown in Table 3, SD modelers did not identify any loop in 25 responses (first row: 23+2) and the SD Bot reported no loop in 23 of those responses. There are two cases in which the SD Bot found a feedback loop in the response but human coders recorded zero loop. There are three cases in which the human coders identified one feedback loop but the SD Bot did not identify the loop. In summary, the feedback identification results of the SD Bot and human SD experts matches in 83% of the cases.

Table 3. Number of feedback loops

|  | SD Bot=0 | SD Bot=1 | SD Bot>1 |
|---|---|---|---|
| Human coders=0 | 23 | 2 | 0 |
| Human coders =1 | 3 | 1 | 0 |
| Human coders >1 | 0 | 0 | 1 |

Note: The human coders rows represent the number of feedback loops coded by the system dynamics modelers, and SD Bot columns represent the number of loops identified by the bot.

## 5. Discussion

This study reports the results of the first computer program, SD Bot, that automates the generation of CLDs from textual data. The bot leverages the recent advances in large language models to identify causal relationships in a text and create a CLD. The performance of this program is evaluated using two datasets, the first one includes 20 textual data and associated CLDs from the system dynamics textbooks and journal publications, and the second one contains the responses of 30 non-SD expert participants to a vignette, and manually coded responses by SD experts. Our evaluation indicates that the SD Bot captures the majority of links and loops of the two evaluation datasets.

Our first evaluation shows that the bot captures 59% of causal relationships and 66% of feedback loops presented. We note that most of the missing links relate to the fact they are not mentioned in the text that described the diagram, probably the authors assuming that they are trivial for the human reader and can be implied from the related figures. For cases that describe all of the links of their diagrams, the bot is often able to recover the CLD close to perfect. We also note that the performance of the SD Bot diminishes with lengthy texts containing less explicit relationships. Future programs can divide the text, identify causal relationships, and then merge the results to improve the outcomes. In the second evaluation dataset, the SD Bot identifies 56% of relationships indicated by the human coders. The number of identified feedback loops by the bot matches the human coders in 83% of cases, however, we should note that in many of these cases there were no feedback loops, and the high rate of correct identification relates to the high true negative rates.

Automating the generation of CLDs using an LLM provides multiple benefits. First, building CLDs using textual data is very labor intensive and such programs can save time. In that sense this paper contributes to studies that have been exploring various ways of turning text to maps (Kim and Andersen 2012). Second, the program can help develop better text-data-informed models and make it easier for novice modelers to develop causal loop diagrams. In addition, as mental models of researchers may impact the generation of CLDs, by comparing human generated CLD with an AI-generated CLD we can check the fidelity of causal loop diagrams and trace them back to the text (Jalali & Beaulieu, 2023). Third, the bot helps systems thinking scholars map and analyze mental models of individuals more efficiently. This has been one of the main challenges of studies analyzing complexity comprehension and systems thinking skills (Haque et al. 2023, Davis et al. 2020). Fourth, the program can be further developed to merge multiple sources and create one CLD that synthesizes different opinions or findings. This can be especially helpful for constructing CLDs through literature reviews. Fifth, such programs can be used in group model building (GMB) sessions to support the modeling team. Conversations of GMB participants can be turned into text and fed to the computer program to support the live mapping of the system. Additionally, the recording of the GMB session can be used after the meeting to identify feedback mechanisms not captured by the facilitator during the session.

The SD Bot created in this project has multiple limitations. First, similar to the other applications of generative AI, the outputs of the SD Bot should be checked by humans. Users should carefully check every relationship and the associated reasoning provided by the SD Bot, as well as the sign of causal links. In the two evaluation datasets consisting of 50 examples, we have not found any error in the polarity of the causal relationships, but an error might occur when a sentence describes many variables and their relationships. Second, the schematic depiction of the generated CLDs are not as clean and organized as those generated by humans. This will require further coding and use of different programming techniques for better visualization. Third, consistency is a concern as the generated CLD may vary slightly when an experiment is repeated, though the change is often minimal.

This project is a first step toward automating the generation of CLDs. This thread of research can substantially improve building system dynamics models. We invite researchers to further develop such computer programs and use our testbed to evaluate their program performance. We also invite practitioners to use the SD Bot and we welcome feedback for further improving it. Instructions to use the bot and the testbed can be found in the supplementary material.

## Supplementary material

Instructions to use the bot can be found at https://github.com/bear96/SDBot-DesktopApp and the testbed can be found at https://github.com/bear96/SDBot-DesktopApp/tree/main/Supplementary%20Data